\documentclass[letter,aps,showpacs,superscriptaddress,nofootinbib,tightenlines,twocolumn,prl]{revtex4}

\usepackage{amsfonts}
\usepackage{multirow}
\usepackage{mathrsfs}
\usepackage{graphicx}
\usepackage{amsmath}
\usepackage{amssymb}
\usepackage{bm}
\usepackage{bbm}
\usepackage{color}

\def\bfsigma{\mbox{\boldmath $\sigma$}}

\def\OMIT#1{}

\newcommand{\nn}{\nonumber}

\newcommand{\beq}{\begin{equation}}
\newcommand{\eeq}{\end{equation}}
\newcommand{\bqa}{\begin{eqnarray}}
\newcommand{\eqa}{\end{eqnarray}}

\begin{document}
\title{\mbox{}\\[10pt]
Next-to-next-to-leading-order QCD corrections to $\chi_{c0,2}\to \gamma\gamma$}


\author{Wen-Long Sang}
 \affiliation{School of Physical Science and Technology, Southwest University, Chongqing 400700, China\vspace{0.2cm}}
 \affiliation{State Key Laboratory of Theoretical Physics, Institute of Theoretical Physics,
 Chinese Academy of Sciences, Beijing 100190, China\vspace{0.2cm}}

\author{Feng Feng}
\affiliation{China University of Mining and Technology, Beijing 100083, China\vspace{0.2cm}}

\author{Yu Jia}
\affiliation{Institute of High Energy Physics and Theoretical Physics Center for
Science Facilities, Chinese Academy of
Sciences, Beijing 100049, China\vspace{0.2cm}}
\affiliation{Center
for High Energy Physics, Peking University, Beijing 100871,
China\vspace{0.2cm}}

\author{Shuang-Ran Liang}
\affiliation{Institute of High Energy Physics and Theoretical Physics Center for
Science Facilities, Chinese Academy of
Sciences, Beijing 100049, China\vspace{0.2cm}}

\date{\today}

\begin{abstract}
We calculate the next-to-next-to-leading-order (NNLO) perturbative corrections to
$P$-wave quarkonia annihilation decay to two photons, in the framework of nonrelativistic QCD (NRQCD) factorization.
The order-$\alpha_s^2$ short-distance coefficients associated with each helicity amplitude
are presented in a semi-analytic form, including the ``light-by-light'' contributions.
With substantial NNLO corrections, we find disquieting discrepancy when confronting our state-of-the-art predictions with the latest \textsf{BESIII} measurements, especially fail to account for the measured $\chi_{c2}\to\gamma\gamma$ width.
Incorporating the effects of spin-dependent forces would even exacerbate the situation, since it lifts
the degeneracy between the nonperturbative NRQCD matrix elements of $\chi_{c0}$ and $\chi_{c2}$
toward the wrong direction.
We also present the order-$\alpha_s^2$ predictions to $\chi_{b0,2}\to\gamma\gamma$,
which await the future experimental test.
\end{abstract}

\pacs{\it 12.38.Bx, 13.20.Gd, 14.40.Pq}


\maketitle

Charmonium decay has historically played an important role in establishing the asymptotic freedom of QCD,
and served as a clean platform to probe the interplay
between pertubative and nonperturbative dynamics~\cite{Appelquist:1974zd,DeRujula:1974nx}.
Among them, the electromagnetic decay $\chi_{c0,2}\to \gamma\gamma$ provide a particularly interesting, and, rich
testing ground of QCD~\cite{Kwong:1987ak,Voloshin:2007dx}.
In the past decades, these decay channels have been extensively studied from
various theoretical angles, such as nonrelativistic potential model~\cite{Li:1990sx,Gupta:1996ak}, relativistic quark model~\cite{Godfrey:1985xj,Munz:1996hb,Ebert:2003mu},
Bethe-Salpeter approach~\cite{Huang:1996bk},
nonrelativistic QCD (NRQCD) factorization~\cite{Bodwin:1994jh,Petrelli:1997ge},
as well as lattice QCD~\cite{Dudek:2006ut}.
On the experimental side, they were previously measured by \textsf{CLEO-c}~\cite{Ecklund:2008hg}.
\textsf{BESIII} experiment~\cite{Ablikim:2012xi} has recently reported
their high precision results,
\begin{subequations}
\bqa
&& \Gamma_{\gamma\gamma}(\chi_{c0})=(2.33\pm 0.20\pm0.13\pm0.17)\:{\rm keV},
\\
&& \Gamma_{\gamma\gamma}(\chi_{c2})=(0.63\pm 0.04\pm0.04\pm0.04)\:{\rm keV}.
\eqa
\label{partial:widths:chic02}
\end{subequations}
In addition, \textsf{BESIII} presents the ratio of the decay rates between
$\chi_{c2}$ and $\chi_{c0}$.
For the first time, they also measured the ratio of the two polarized
decay rates for $\chi_{c2}$:
\begin{subequations}
\label{experiment:BESIII}
\bqa
&& {\mathcal R} =\frac{\Gamma_{\gamma\gamma}(\chi_{c2})}{\Gamma_{\gamma\gamma}(\chi_{c0})}
=0.271\pm0.029\pm0.013\pm0.027,\qquad
\label{BES3:measurement:R:ratio}
\\
&& f_{0/2}=\frac{\Gamma^{\lambda=0}_{\gamma\gamma}(\chi_{c2})}
{\Gamma^{\lambda=2}_{\gamma\gamma}(\chi_{c2})}
=0.00\pm0.02\pm0.02,
\label{BES3:measurement:f02:ratio}
\eqa
\end{subequations}
where $\lambda=|\lambda_1-\lambda_2|$, and $\lambda_1, \lambda_2=\pm 1$
denote the helicities of the outgoing photons.
The precise data thereby calls for the full-fledged
theoretical inspection.

In parallel with positronium decay, the leading-order NRQCD prediction to
$\chi_{c0,2}\to\gamma\gamma$ in the nonrelativistic limit is extremely simple,
yields ${\mathcal R}=4/15\approx0.27$~\cite{Barbieri:1975am}.
This is impressively consistent with the measurement
(\ref{BES3:measurement:R:ratio}).
Nevertheless, these processes are sensitive to the next-to-leading-order (NLO)
radiative correction~\cite{Barbieri:1980yp,Barbieri:1981xz},
with the predicted ${\mathcal R}$ scattered in the range from
0.09 to 0.36~\cite{Gupta:1996ak,Godfrey:1985xj}.

To date, the next-to-next-to-leading-order (NNLO) radiative corrections
are only available for a few $S$-wave quarkonium electromagnetic
decay processes, exemplified by $\Upsilon(J/\psi)\to e^+e^-$~\cite{Czarnecki:1997vz,Beneke:1997jm}, $\eta_{b,c}\to\gamma\gamma$~\cite{Czarnecki:2001zc,Feng:2015uha},
and $B_c\to \ell \nu$~\cite{Onishchenko:2003ui,Chen:2015csa}, as well as the $\gamma\gamma^*\to \eta_{c,b}$
transition form factor~\cite{Feng:2015uha}.
It has been found that the
NNLO radiative corrections in aforementioned processes are often negative and substantial.
The goal of this work is to address the complete
NNLO corrections to $P$-wave quarkonium annihilation into two photons.

The partial widths for $\chi_{c0,2}\to\gamma\gamma$ can be expressed as
\begin{subequations}
\label{partial:width:helicity:ampl}
\bqa
 \Gamma_{\gamma\gamma}(\chi_0) &=&  {1 \over 16\pi}
 \bigg(2 |{\cal A}^{\chi_0}_{1,1}|^2\bigg),
 \\
  \Gamma_{\gamma\gamma}(\chi_2)&=&\frac{1}{5}{1\over 16\pi}
 \bigg(2|{\cal A}^{\chi_2}_{1,1}|^2+2|{\cal A}^{\chi_2}_{1,-1}|^2\bigg),
\eqa
\end{subequations}
where ${\cal A}^{\chi_J}_{\lambda_1,\lambda_2}$ signifies the helicity amplitude
for $\chi_{cJ}\to \gamma(\lambda_1)\gamma(\lambda_2)$.
We have employed parity invariance to only enumerate
the independent helicity amplitudes
in (\ref{partial:width:helicity:ampl}).

NRQCD factorization approach, which exploits the nonrelativistic nature of heavy quarkonium,
provides a systematic and model-independent framework to tackle quarkonium decay~\cite{Bodwin:1994jh}.
At the lowest order in $v$, the helicity amplitudes in (\ref{partial:width:helicity:ampl})
can be written in a factorized form:
\bqa
\label{NRQCD:factorization:chic:EM:decay}
 &&  {\cal A}_{\lambda_1,\lambda_2}^{\chi_{0,2}} =  {\mathcal C}_{\lambda}^{\chi_{0,2}}(m, \mu_R,\mu_\Lambda)\,
 {{\langle 0 \vert \chi^\dagger {\cal K}_{^3P_{0,2}}\psi(\mu_\Lambda)\vert \chi_{c0,2} \rangle}\over m^{3/2} }
\nn\\
&& \qquad +{\mathcal O}(v^2),
\eqa
where
\begin{subequations} \label{NRQCD:operator}
\bqa
{\cal K}_{^3P_{0}}&=&\frac{1}{\sqrt{3}}(-\frac{i}{2}\tensor{{\bf
D}}\cdot\bfsigma),\\
{\cal K}_{^3P_2}&=&-\frac{i}{2}\tensor{D}^{(i}\sigma^{j)}\epsilon^{*ij},
\eqa
\end{subequations}
with $\epsilon^{ij}$ representing the polarization tensor of $\chi_{c2}$.

${\mathcal C}_{\lambda}^{\chi_{0,2}}(m, \mu_R,\mu_\Lambda)$  in
(\ref{NRQCD:factorization:chic:EM:decay})
signifies the NRQCD short-distance coefficient (SDC),
where $m$, $\mu_R$,  $\mu_\Lambda$ denote the charm quark mass, renormalization scale,
and NRQCD factorization scale, respectively.
In phenomenological analysis, these nonpertubative matrix elements are often substituted as
the derivative of $P$-wave radial Schr\"{o}dinger wave functions at
the origin:
\bqa
\langle 0 \vert \chi^\dagger {\cal K}_{^3P_{0,2}}\psi(\mu_\Lambda)\vert \chi_{c0,2} \rangle &=&
\sqrt{3N_c\over 2\pi} \overline{R_{\chi_{c0,2}}^\prime}(\mu_\Lambda),
\label{matrix:element:wavefunction:at:origin}
\eqa
where $N_c=3$ is the number of color. In literature, it is often assumed that
$\overline{R_{\chi_{c0}}^\prime} \approx \overline{R_{\chi_{c2}}^\prime}$ by invoking the
approximate heavy quark spin symmetry (HQSS).
We stress that, in NRQCD these wave functions at the origin are promoted as
scale-dependent quantities.

Thanks to the asymptotic freedom, the SDCs can be computed order by order in $\alpha_s$.
Through NNLO in $\alpha_s$, the SDC affiliated with the only helicity channel of
$\chi_{c0}$ is
\bqa
\label{A0:LO}
{\cal C}^{\chi_0}_{0} &=& \frac{4\sqrt{3}\pi e_Q^2\alpha}{\sqrt{m}}
\Bigg\{1+C_F {\alpha_s(\mu_R)\over \pi}\bigg(\frac{\pi^2}{8}
-\frac{7}{6}\bigg)
\nn\\
&+& \frac{\alpha_s^2}{\pi^2}\bigg[C_F\frac{\beta_0}{4}\bigg(\frac{\pi^2}{8}-\frac{7}{6}\bigg)
\ln\frac{\mu_R^2}{m^2}+\Delta^{\chi_0}_{0}\bigg]\Bigg\},
\eqa
and two independent SDCs ${\mathcal C}_{0,2}^{\chi_{2}}$ are
\begin{subequations}
\label{A2:LO}
\bqa \label{SDC:chi2:lambda:0}
 {\mathcal C}^{\chi_2}_{0} &=& \frac{4\sqrt{6}\pi\alpha e_Q^2}{3\sqrt{m}}
\Bigg\{C_F\frac{\alpha_s(\mu_R)}{\pi}\bigg(\frac{3\pi^2}{8}-6\ln2+1\bigg)
\\
&+& \frac{\alpha_s^2}{\pi^2} \bigg[C_F\frac{\beta_0}{4}\bigg(\frac{3\pi^2}{8}-6\ln2+1\bigg)\ln\frac{\mu_R^2}{m^2}+
\Delta^{\chi_2}_0\bigg]\Bigg\},
\nn\\
  {\mathcal C}^{\chi_2}_{2}&=& -\frac{8\pi\alpha e_Q^2}{\sqrt{m}}\Bigg\{ 1-2C_F\frac{\alpha_s(\mu_R)}{\pi}
\nn\\
& +&  \frac{\alpha_s^2}{\pi^2} \bigg(-2C_F\frac{\beta_0}{4}\ln\frac{\mu_R^2}{m^2}+\Delta^{\chi_2}_2\bigg)\Bigg\}.
\eqa
\end{subequations}
$\beta_0 = {11\over 3}C_A - {2\over 3}(n_L+n_H)$ is
the one-loop coefficient of the QCD $\beta$-function, where
$n_H=1$, and $n_L$ signifies the number of light quark flavors ($n_L=3$ for $\chi_c$, $4$ for $\chi_b$).
The occurrence of the $\beta_0\ln \mu_R$ term in (\ref{A0:LO}) and (\ref{A2:LO})
is demanded by renormalization group invariance.

To the best of our knowledge, the NLO perturbative correction to the $\lambda=0$ amplitude in (\ref{SDC:chi2:lambda:0})
is new. Interestingly, this helicity amplitude turns out to vanish at Born level.
Thus, NRQCD framework appears to offer a natural explanation for the tiny value of $f_{0/2}$ in
(\ref{BES3:measurement:f02:ratio}) observed by \textsc{BESIII}.

The nontrivial task is then to decipher $\Delta^{\chi_{0,2}}_{0,2}$.

Rather than follow the literal matching doctrine, we employ the standard shortcut of directly extracting the SDCs~\cite{Czarnecki:1997vz,Beneke:1997jm}.
We compute the on-shell quark amplitude for $c\bar{c}({}^3 P_J^{(1)})\to \gamma\gamma$ through order $\alpha_s^2$.
In contrast to the $S$-wave quarkonium decay, we expand the corresponding amplitude to the first order in $q$, the relative momentum between $c$ and $\bar{c}$, to identify the $P$-wave component,
and compose the $c\bar{c}({}^3 P_{0,2}^{(1)})$ state via the standard procedure~\cite{Petrelli:1997ge}.
In the end we project out the respective helicity amplitudes.
A key simplification originates from the fact that, when conducting the loop integration,
$q$ has already been set to zero.


\begin{figure}[t]
\centering
\includegraphics[width=0.48\textwidth]{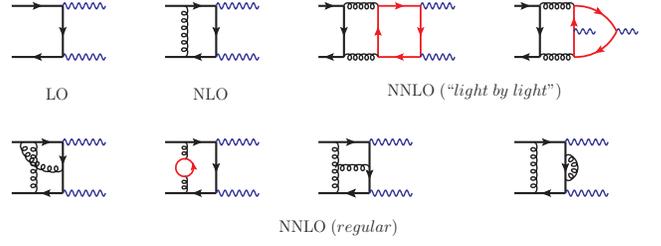}
\caption{Representative Feynman diagrams for $c\bar{c}({}^3 P_J^{(1)})\to \gamma\gamma$
through order $\alpha_s^2$.
\label{feyn:diag}}
\end{figure}

We briefly describe the calculation.
The package \textsf{FeynArts}~\cite{Hahn:2000kx} is employed to generate corresponding Feynman diagrams and
amplitudes through ${\cal O}(\alpha_s^2)$ in Feynman gauge.
There are 108 regular 2-loop diagrams and
12 ``light-by-light" (LBL) scattering diagrams, some of which are sketched in Fig.~\ref{feyn:diag}.
The latter gauge-invariant subsets are UV- and IR-finite.
Dimensional regularization (DR) is employed to regularize both UV and IR divergences.
We then use \textsf{FeynCalc/FormLink}~\cite{Mertig:1990an,Feng:2012tk} to carry out the trace over Dirac/color matrices.
The packages \textsf{Apart}~\cite{Feng:2012iq} and \textsf{FIRE}~\cite{Smirnov:2014hma} are utilized
to conduct partial fraction together with integration-by-parts (IBP) reduction.
Finally, we end up with around 80 master integrals (MI).
For a dozen of simpler MIs, we employ the $\alpha$ parameters~\cite{Smirnov:2012gma} as well as
the Mellin-Barnes tools \textsf{AMBRE}~\cite{Gluza:2007rt}/\textsf{MB}~\cite{Czakon:2005rk} to
infer the (semi-) analytic expressions; for the multi-leg two-loop MIs,
we combine \textsf{FIESTA}/\textsf{CubPack}~\cite{Smirnov:2013eza,CubPack} to carry out
sector decomposition and subsequent numerical integrations with quadruple precision.

The order-$\alpha_s^2$ expressions for the heavy quark wave function and mass renormalization constants are taken from \cite{Broadhurst:1991fy, Melnikov:2000zc}. The strong coupling constant is renormalized to one-loop order
under $\overline{\rm MS}$ scheme. All the UV divergences are eliminated by the renormalization procedure.
However, at this stage, the NNLO amplitudes still
contain single IR poles, with coefficients differing from the ${}^3P_{0}$  to the ${}^3P_{2}$ channel.

These single IR poles are intimately connected to the anomalous dimensions of the NRQCD color-singlet currents associated with ${}^3P_{0,2}$, as specified in (\ref{NRQCD:operator}).
In fact, from the lower-energy effective field theory of NRQCD, Hoang and Ruiz-Femenia are able to predict
the anomalous dimensions for NRQCD bilinear carrying general ${}^{2S+1}L_J$ quantum number~\cite{Hoang:2006ty}.
Particulary, the anomalous dimensions of the operators carrying quantum number ${}^3P_{J}$
are predicted to be
\begin{subequations}\label{AD:3PJ:currents}
\bqa
&&\gamma_{{}^3P_0}= -C_F\left({C_F\over 6}+ {C_A\over 24} \right) \alpha_s^2+{\cal O}(\alpha_s^3),
\\
&&\gamma_{{}^3P_2}= -C_F\left( {13 C_F\over 240} + {C_A \over 24} \right)\alpha_s^2+{\cal O}(\alpha_s^3).
\eqa
\end{subequations}
Their difference signals the violation of HQSS due to spin-dependent interactions.

It is reassuring that the coefficients of the uncancelled IR poles in our NNLO amplitudes
turn out to exactly match the UV poles as implied in (\ref{AD:3PJ:currents}).
In our opinion, this is a highly nontrivial verification of the correctness of our calculation.

We thereby factorize these IR poles into the corresponding $\chi_{c0,2}$-to-vacuum
NRQCD matrix elements in (\ref{NRQCD:factorization:chic:EM:decay})
under $\overline{\rm MS}$ prescription,
with $\ln \mu_\Lambda$ now manifested in the respective SDCs.

The $\Delta^{\chi_J}_\lambda$ receives contributions from both regular and LBL diagrams,
where the former is real valued, and the latter complex valued.
It is convenient to decompose $\Delta^{\chi_J}_\lambda$ into two parts:
\bqa\label{Delta:decomposition}
\Delta^{\chi_J}_{\lambda}= \Delta^{\chi_J}_{{\rm reg},\:\lambda}
+ \Delta^{\chi_J}_{{\rm lbl},\:\lambda}.
\eqa
The regular part can be organized according to their color structure:
\bqa
\label{Delta:decomp:color:structure}
\Delta^{\chi_J}_{{\rm reg},\:\lambda}  &=&
C_F^2 s_{A;\lambda}^{\chi_J}+ C_F C_A s_{NA,\lambda}^{\chi_J}+ n_L C_F T_F s_{L,\lambda}^{\chi_J}
\nn\\
&+& n_H C_F T_F s_{H,\lambda}^{\chi_J},
\eqa
where $C_F=\tfrac{4}{3}$, $C_A=3$, $T_F=\tfrac{1}{2}$ are $SU(3)$ color factors.

The regular pieces of the only helicity component for $\chi_0$ are
\bqa
\label{Delta0:reg}
s_{A,0}^{\chi_0}&=&-\frac{2\pi^2}{3}\ln\frac{\mu_\Lambda}{m}-9.14751077(6),
\nn\\
s_{NA,0}^{\chi_0}&=&-\frac{\pi^2}{6}\ln\frac{\mu_\Lambda}{m}-1.69821088(5),
\nn\\
s_{L,0}^{\chi_0}&=&\frac{1}{432}\bigg[-126 \zeta(3)-45\pi^2+244\bigg],
\nn\\
s_{H,0}^{\chi_0}&=&0.09292479(2).
\eqa

The regular pieces affiliated with the two helicity components of $\chi_2$ are
\bqa\label{Delta2-0:reg}
s_{A,0}^{\chi_2}&=&-1.59023228(9),\nn\\
s_{NA,0}^{\chi_2}&=&2.13274690(5),\nn\\
s_{L,0}^{\chi_2}&=&-\frac{7}{8}\zeta(3)-\frac{3\pi^2}{16}-2\ln^2 2+\frac{16}{3}\ln2-\frac{5}{9},\nn\\
s_{H,0}^{\chi_2}&=&0.01594186(1),
\eqa
and
\bqa\label{Delta2-2:reg}
s_{A,2}^{\chi_2} &=&-\frac{13\pi^2}{60}\ln\frac{\mu_\Lambda}{m}-5.93023533(7),\nn\\
s_{NA,2}^{\chi_2} &=&-\frac{\pi^2}{6}\ln\frac{\mu_\Lambda}{m}-5.78204922(4),\nn\\
s_{L,2}^{\chi_2} &=&\frac{43}{36}+\frac{\pi^2}{16},\nn\\
s_{H,2}^{\chi_2} &=&0.021716502(9).
\eqa
Note that the absence of $\ln\mu_\Lambda$ in (\ref{Delta2-0:reg}) originates from
the vanishing of LO amplitude for
the helicity configuration $\chi_2\to\gamma(\pm 1)\gamma(\pm 1)$.

In contrast to regular part, it is rather challenging for \textsf{FIESTA} to
acquire high-precision results for the complex-valued MIs associated with
the LBL diagrams. Fortunately, some of them can be worked out analytically.
Employing the $\alpha$-parameters~\cite{Smirnov:2012gma} or Mellin-Barnes tools~\cite{Czakon:2005rk,Gluza:2007rt},
it is always feasible to reduce the remaining MIs into one or two-dimensional integrals,
which can then be readily
computed with high numerical precision.

The LBL part for the $\chi_0\to\gamma(\pm 1)\gamma(\pm 1)$ is
\bqa\label{AD:3P0}
&& \Delta^{\chi_0}_{{\rm lbl},0} = (-0.120326+0.398547i)n_H C_F T_F
\nn\\
&&+\bigg(0.953741+{i\pi \over 6}\bigg) C_F T_F \sum_{i}^{n_L}{e_i^2\over e_Q^2},
\eqa
where $e_i$ represents the electric charge of the $i$-th light flavor.

The LBL pieces associated with the two helicity components of $\chi_2$ are
\begin{subequations}\label{Delta2:lbl}
\bqa
&&\Delta^{\chi_2}_{{\rm lbl},0}=\bigg(-0.019772+0.011196\, i\bigg)n_H C_F T_F
\nn\\
&&+\bigg[0.359850+i\pi\left(\frac{7\pi^2}{6}-{23\over 2}\right)\bigg] C_F T_F \sum_{i}^{n_L} {e_i^2\over e_Q^2},\nn\\
\\
&&\Delta^{\chi_2}_{{\rm lbl},2}=\bigg(-0.088227+0.187239\, i\bigg)n_H C_F T_F
\nn\\
&&+\bigg[-0.669873+\frac{\pi}{27}(91-12\pi^2+24\ln2)i\bigg]
\nn\\
 &&\times C_F T_F \sum_{i}^{n_L}\frac{e_i^2}{e_Q^2}.
\eqa
\end{subequations}
In passing, we recall that the rare decay process $\chi_{c2}\to e^+e^-$
contains uncancelled IR divergences~\cite{Kuhn:1979bb}. Since the occurring one-loop box diagrams just comprise
subdiagrams of
our two-loop LBL diagrams, it is intriguing that our LBL contributions turn out to be completely IR finite.

With all the order-$\alpha_s^2$ terms in (\ref{Delta:decomposition}) available in a semi-analytic form,
we can assemble them together to deduce the corresponding SDCs in (\ref{A0:LO}), (\ref{A2:LO}), and substitute them into (\ref{NRQCD:factorization:chic:EM:decay}) to deduce the respective helicity amplitudes, finally obtain the desired two-photon widths for $\chi_{c0,2}$ according to (\ref{partial:width:helicity:ampl}).

First we would like to predict $f_{0/2}$ and ${\mathcal R}$ and compare with \textsf{BESIII} experiments.
Following the analysis conducted for the ratio of the decay rates of $J/\psi\to e^+e^-$ to $\eta_c\to\gamma\gamma$~\cite{Czarnecki:2001zc,Kiyo:2010jm},
we also expand these two ratios strictly to the second order in $\alpha_s$:
\begin{subequations}\label{NRQCD:NNLO:f:R}
\bqa
\label{R:scheme2}
{\mathcal R} &=& {4\over 15}\Omega \bigg[1-\bigg({\pi^2\over 3}+{20\over 9}\bigg){\alpha_s\over \pi}
\\
&-& \bigg(5.855 +22.967 \,\ln{\mu_R\over m}
+15.791\,\ln {m\over \mu_\Lambda}\bigg)\left({\alpha_s\over \pi}\right)^2 \bigg],
\nn\\
f_{0/2} &=& {\alpha_s^2\over 216\pi^2}\left(8+3\pi^2-48
\ln 2 \right)^2,
\eqa
\end{subequations}
where $n_L=3$ has been taken, and $\Omega$ is defined by
\beq
\Omega =\bigg\vert {\overline{R_{\chi_{c2}}^{\prime}} (\mu_\Lambda)\over \overline{R_{\chi_{c0}}^{\prime}}(\mu_\Lambda)}\bigg|^2,
\label{def:Omega}
\eeq
which characterizes the extent of the violation of HQSS.

With the nonperturbative matrix elements cancelled, the helicity ratio $f_{0/2}$ is entirely
determined by the order-$\alpha_s$ (leading) contribution of the $\lambda=0$ component from $\chi_2$ decay
in (\ref{SDC:chi2:lambda:0}).

In the following phenomenological analysis, we will use
the two-loop quark pole masses as $m_c=1.68$ GeV and $m_b=4.78$ GeV~\cite{Feng:2015uha}.
Running strong coupling at a given scale is evaluated by the package
\textsf{RunDec}~\cite{Chetyrkin:2000yt}.

We first present the NRQCD predictions accurate to NLO in $\alpha_s$:
\bqa\label{fR-NLO}
{\mathcal R}=(0.124_{-0.028}^{+0.032})\,\Omega,\qquad f_{0/2}=0,
\eqa
where the uncertainty comes from varying the renormalization scale
in the range $1\;{\rm GeV}<\mu_R<2m$, with central value at $\mu_R = m$.
Assuming $\Omega=1$, the predicted ${\cal R}$ then becomes considerably smaller than 
the \textsf{BESIII} data in (\ref{BES3:measurement:R:ratio}).

From (\ref{NRQCD:NNLO:f:R}), we further give our predictions at
NNLO accuracy:
\bqa
{\mathcal R}=(0.075_{-0.051}^{+0.044})\,\Omega,
\qquad f_{0/2}=0.0009_{-0.0004}^{+0.0009},
\label{fR-NNLO}
\eqa
with the central values obtained by setting $\mu_\Lambda=1$ GeV and $\mu_R=m$.
Two kinds of uncertainties are included by sliding the $\mu_\Lambda, \mu_R$ in the range $\tfrac{m}{2}<\mu_\Lambda<m$ and $1\,{\rm GeV}<\mu_R<2m$, respectively.

While the very tiny $f_{0/2}$ predicted in (\ref{fR-NNLO}) fully agrees with the \textsf{BESIII} measurement within errors,
the NNLO prediction of ${\cal R}$ deviates further from the data
relative to the NLO prediction, in the HQSS limit.

If the HQSS-violating effects would lead to $\Omega>1$, our NNLO predictions in (\ref{fR-NNLO}) would still have chance
to agree with the data.
The spin-dependent interactions such as spin-orbital force and tensor force have been incorporated to study the fine splitting among $\chi_{cJ}$~\cite{Brambilla:2004wu}. In order to elucidate the role played by the HQSS violation,
we have implemented these spin-dependent forces within the Cornell potential model 
[$\chi_{c0(2)}$ acquires a repulsive (attractive) ${1\over r^3}$ potential, respectively], and
found that the curvatures of the radial wave functions of $\chi_{c0}$ and $\chi_{c2}$
are changed towards the opposite direction such that $\Omega < 1$. 
Therefore, the discrepancy between (\ref{fR-NNLO}) and the \textsf{BESIII}
measurement of $\cal R$ even further deteriorates!

\setlength{\tabcolsep}{5pt}
\begin{table}
\begin{tabular}{ccccc}
\hline\hline
 & $\mu_\Lambda$  & LO & NLO & NNLO \cr
\hline
\multirow{2}{*}{$\chi_{c0}$}
& $1$ GeV& \multirow{2}{*}{$0.032\pm0.004$} & \multirow{2}{*}{$0.036\pm0.005$}
& $0.091_{-0.024}^{+0.087}$ \cr
& $m$&  & & $0.127_{-0.049}^{+8.598}$ \cr
\hline
\multirow{2}{*}{$\chi_{c2}$}
& $1$ GeV& \multirow{2}{*}{$0.032\pm0.004$} & \multirow{2}{*}{$0.076_{-0.021}^{+0.031}$}
& -- \cr
& $m$&  &  & -- \cr
\hline\hline
\end{tabular}
\caption{Determination of $|\overline{R_{\chi_{cJ}}^{\prime}} (\mu_\Lambda)|^2$ (${\rm GeV}^5$) from \textsf{BESIII} data 
at various level of perturbative accuracy.
The uncertainty is estimated by combining the experimental error with that by varying $\mu_R$
from $1$ GeV to $2m$.
\label{Table:1}}
\end{table}

To closely examine the impact of NNLO corrections, we can also extract the nonperturbative factors 
$\overline{R_{\chi_{c0,2}}^\prime} (\mu_\Lambda)$
from the measured two-photon widths of $\chi_{c0,2}$ in (\ref{partial:widths:chic02}).
In Table~\ref{Table:1} we tabulated these fitted parameters at various levels of accuracy in $\alpha_s$.
Although the NNLO corrections to $\chi_{c0}\to\gamma\gamma$ are sizable,
one is still able to obtain a reasonable value for the matrix element; however,
for the $\chi_{c2}\to\gamma\gamma$, both NLO and NNLO corrections are negative yet substantial,
such that the partial width turns negative in some parameter space, and we refrain from
listing the fitted value of $\overline{R_{\chi_{c2}}^\prime}$ in
Table~\ref{Table:1}.

\begin{figure}[tb]
\centering
\includegraphics[width=0.5\textwidth]{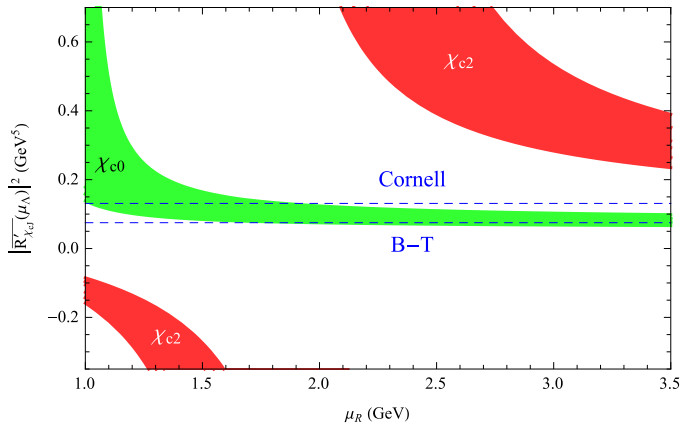}
\caption{The dependence of
$|\overline{R_{\chi_{cJ}}^{\prime}} (\mu_\Lambda)|^2$, which are fitted from the \textsf{BESIII} data using the NNLO formula,
as a function of $\mu_R$.
The blue and green bands are obtained by varying $\mu_\Lambda$ from 1 GeV to $m$,
and the two horizontal lines correspond to the respective values given by B-T and Cornell potential models~\cite{Eichten:1995ch}.
\label{fig:O1}}
\end{figure}

In Fig.~\ref{fig:O1}, we show the values of $|\overline{R_{\chi_{c0,2}}^{\prime}} (\mu_\Lambda)|^2$ 
fitted to account for the \textsf{BESIII} data following the NNLO formula,
as a function of $\mu_R$.
For $\chi_{c0}\to\gamma\gamma$, within reasonable choice of $\mu_R$ and $\mu_\Lambda$, the fitted
$|\overline{R_{\chi_{c0}}^{\prime}}|^2$ agrees with those predicted from the famous Cornell and
Buchm\"{u}ller-Tye (B-T)
potential models~\cite{Eichten:1995ch}.
However, for small $\mu_R$, the fitted $|\overline{R_{\chi_{c2}}^{\prime}}|^2$ becomes negative, hence unphysical;
for large $\mu_R$, $|\overline{R_{\chi_{c2}}^{\prime}}|^2>|\overline{R_{\chi_{c0}}^{\prime}}|^2$ so that $\Omega >1$,
in contradiction to what is implied by the spin-dependent force.
While the NNLO corrections to $\chi_{c0}\to\gamma\gamma$ are under theoretical control,
it appears rather challenging to account for the $\chi_{c2}\to\gamma\gamma$ data
from our results.

It is straightforward to adapt (\ref{NRQCD:NNLO:f:R}) to analyze $P$-wave bottomonia decays to two photons,
by taking $n_L=4$. The NLO perturbative predictions are
\bqa\label{fR-b-NLO}
{\mathcal R}^b=(0.169_{-0.073}^{+0.015})\,\Omega^b,\qquad f^b_{0/2}=0,
\eqa
where $\Omega^b$ is the bottomonium counterpart of (\ref{def:Omega}). 
After incorporating the NNLO corrections, we then predict
\bqa\label{fR-b-NNLO}
{\mathcal R}^b=(0.126^{+0.025}_{-0.144})\,\Omega,\qquad f^b_{0/2}=0.0004^{+0.0014}_{-0.0001},
\eqa
where the central values are obtained by setting $\mu_\Lambda=\tfrac{m_b}{2}$ and $\mu_R=m_b$.
The uncertainty is estimated by varying $\mu_\Lambda, \mu_R$ in the range $1\,{\rm GeV}<\mu_\Lambda<m_b$  and $1\,{\rm GeV}<\mu_R<2m_b$.
Notably, the convergence of perturbative expansion for $\chi_{bJ}\to \gamma\gamma$ has been considerably improved
with respect to $\chi_{cJ}$ decay, and we also expect here the HQSS-violation has smaller impact.

To summarize, in this work we have computed, for the first time,
the complete order-$\alpha_s^2$ corrections to $\chi_{c,b}\to \gamma\gamma$ in NRQCD framework,
deducing the corresponding SDCs for each helicity amplitude.
The NNLO corrections to $\chi_{c0,2}\to \gamma\gamma$ are found to be substantial,
and we find it rather difficult to account for the ratio of
their decay rates recently measured by \textsf{BESIII}.
This discrepancy even deteriorates after incorporating the spin-dependent inter-quark interaction.
To resolve this puzzle, it is worth computing higher-order radiative corrections,
as well as including the relativistic corrections. 
However, our poor knowledge of the higher-order $P$-wave NRQCD matrix elements
renders a sharp order-$v^2$ prediction unrealistic~\cite{Ma:2002eva,Brambilla:2006ph} .
In contrast, we believe our ${\cal O}(\alpha_s^2)$ predictions 
to $\chi_{b0,2}\to \gamma\gamma$ are trustworthy. 
Hopefully, the forthcoming \textsf{Belle II} experiments, 
and the next-generation high-energy colliders, 
will have a bright prospect to measure these two-photon decay channels, 
thereby test our predictions.

\begin{acknowledgments}
{\noindent\it Acknowledgment.}
We thank Estia Eichten for enlightening discussion on the spin-dependent inter-quark force.
W.-L.~S. is supported by the National Natural Science Foundation of China under Grant No.~11447031, and
by the Open  Project Program of State Key Laboratory of Theoretical Physics under Grant No.~Y4KF081CJ1,
and also by the Fundamental Research Funds for the Central Universities under Grant No. SWU114003, No. XDJK2016C067.
The work of F.~F. is supported by the National Natural Science Foundation of China under Grant No.~11505285,
and by the Fundamental Research Funds for the Central Universities.
The work of Y.~J. and S.-R.~L. is supported in part by the National Natural Science Foundation of China under Grants
No.~11475188, No.~11261130311 (CRC110 by DGF and NSFC), by the IHEP Innovation Grant under contract number Y4545170Y2,
and by the State Key Lab for Electronics and Particle Detectors.
\end{acknowledgments}

\end{document}